# Broadband Excitation and Active Control of Terahertz Plasmons in Graphene

Mohammad Ali Khorrami, Samir El-Ghazaly

Department of Electrical Engineering, University of Arkansas, Faytteville, AR

***Abstract*—A novel broadband technique to effectively launch plasmons along a single graphene layer at terahertz (THz) frequencies is proposed. To this end, the coupling of the electromagnetic wave from a readily available plasmonic waveguide established by a periodically corrugated metallic surface to the graphene sheet is proposed. As will be shown, this technique can significantly surmount the need for efficient excitation of plasmons in graphene. For this purpose, an analytical technique based on transmission line theory is employed to calculate the scattering parameters of the connection of the plasmonic waveguides. In this manner, the gating effects of the graphene waveguide on the input reflection and transmission of the junction are also investigated. For comparison, a full wave numerical simulator is employed.**

## I. INTRODUCTION

Graphene is a carbon-based, two-dimensional (2D) nanomaterial revealed less than a decade ago [1]. Afterwards, graphene has been extensively examined as a platform for future photonic and electronic devices. This is due to its extremely high carrier mobility at room and cryogenic temperatures (up to 230,000 $cm^2$ / Vs for suspended exfoliated graphene at T = 5 K [2]), as well as exceptional thermal and mechanical properties [3]. Additionally, the charge density and the surface conductivity of a graphene sheet can be effectively controlled by applying a perpendicular electric field. Moreover, surface waves coupled to carriers, mostly called plasmons can propagate distances up to 100 wavelengths along graphene layers with negligible attenuations in upper section of terahertz (THz) frequency range [4]. All these unique properties have made graphene a promising platform for future compact active plasmonic devices and systems [4]. Plasmonic structures implemented inside two-dimensional electron gas layers of hetero-structures [5]-[8] and graphene [9]-[10] have been vastly explored to develop compact terahertz sources and detectors. In addition, the emergence of plasmon-based logic gates [11] has introduced another beyond CMOS technology alternative that once combined with some of today's best logic design paradigms and practices [12]-[13] may revolutionize the future of computing. In spite of the numerous prospective applications of graphene-based structures, the key remaining challenge is how to efficiently excite the plasmons in graphene using an incident radiating mode electromagnetic wave. This problem is originated from the large phase mismatch between the incident and the plasmonic waves.

Recently, a near field scattering setup with an atomic force microscopy tip and infrared excitation light has been employed to launch plasmons along a graphene layer [14]. However, this is very inefficient technique with very negligible percentage of incident field coupled to the surface wave [15]. Moreover, surface acoustic waves [16] and modulated graphene conductivity [17] have been suggested to launch graphene plasmons. Subsequently, plasmon excitation along a graphene sheet laid over a fabricated silicon diffractive grating and illuminated by an incident radiating mode EM field has been achieved in [15]. Unfortunately, the suggested technique is successful only in a single wave number which is related to the silicon grating period. Therefore, the urgent need to effectively launch the propagating plasmonic mode along graphene layers in a wide frequency range still exists.

In this paper, the possibility of launching the plasmons along a suspended graphene sheet using another plasmonic waveguide is investigated. As will be shown, the surface wave on the interface of a corrugated metal and a dielectric can appropriately launch plasmons along a suspended graphene layer at terahertz frequencies. These specific surface waves traveling on the exterior of a metal with engineered cuts and grooves are mostly called Spoof Surface Plasmon Polaritons (SSPPs) [18]. The SSPPs can be effectively launched using a network analyzer source or a quantum cascade laser. Here, the transmission and the input reflection of the plasmonic wave traveling from the indented metallic structure to a suspended graphene sheet are considered in a wide frequency range. To report and compare the results, conventional microwave theory scattering parameter notation is used.

This paper is organized as follows. In section II, the details of the proposed mechanism to launch plasmons along a graphene layer and the simulation approaches are discussed. Next, the scattering parameters obtained from the full wave simulation of the junction between the waveguides and the analytical model are compared in section III.

## II. SIMULATION DETAILS

Fig. 1.(a) depicts a metallic surface corrugated with linearly spaced grooves with period D, distance (D - A), height H, filled with a dielectric (air) with permittivity $\varepsilon_{r\text{-}SSPP} = 1$. On the last edge of the indented metal, a graphene layer is located (see Fig. 1.(a)). Here, the details of the analytical and full wave simulation of the structure in Fig. 1. (a) are described.

*A. 2D Plasmons along Graphene*

The surface conductivity of the graphene layer $\sigma_g$ is mostly calculated by Kubo formalism [4]. Using the computed conductivity, it can be proved that a $TM^x$ mode electromagnetic wave (known as 2D plasmons) may propagate along a graphene sheet as $|E_F| > \bar{h} \times \omega$ [4], where $E_F$, $\bar{h}$ and $\omega = 2\pi \times f$ are graphene Fermi energy, reduced Planck constant and radial frequency, respectively. The condition is easily satisfied in terahertz and infrared frequency range. Furthermore, the Fermi energy can be altered by applying perpendicular electric fields $E_0$ (see Fig. 1 (a)). The surface wave field variations follow $\exp\left(j\omega t - \gamma_G x - \delta_{G1,2} y\right)$, where $\delta_{G1} = \delta_G$ as $y \geq 0$ and $\delta_{G2} = -\delta_G$ if $y < 0$. Besides, $\gamma_G = \alpha_G + j \times \beta_G$, $\alpha_G$ and $\beta_G$ are the 2D plasmon propagation, attenuation and phase constants, respectively. After solving Maxwell equations and applying boundary conditions, the dispersion relation of the 2D plasmons is obtained:

$$\gamma_G = \pm\sqrt{\frac{4\omega^2\varepsilon_G{}^2}{\sigma_G{}^2} - \frac{\omega^2\varepsilon_{r-G}}{c^2}} \qquad (1)$$

where, $\varepsilon_G = \varepsilon_0 \times \varepsilon_{r\text{-}G}$ ($\varepsilon_0 = 8.85 \times 10^{-12}$ F / m) and $c = 3 \times 10^8$ $(\mathrm{m/s})$ [15]. Moreover, the graphene characteristic impedance is chosen similar to the $TM^x$ mode wave impedance $Z_G = \gamma_G / (j\omega \times \varepsilon_G)$. Here, the measured transport parameters of the suspended graphene layer (in air $\varepsilon_{r\text{-}G} = 1$) with extremely high electron mobility $\mu$ = 230,000 $cm^2$ $V^{-1}$ $s^{-1}$ at T = 5K is considered [2].

*B. Spoof Surface Plasmon Polaritons along a Corrogated Metall*

As proved in [19], the dispersion relation of the fundamental plasmonic mode (with components $E_x$, $E_y$ and $H_z$) that can propagate along an indented perfect electric conductor filled with a dielectric (with relative permittivity $\varepsilon_{r\text{-}SSPP}$) is:

$$\frac{\sqrt{\beta_{SSPP}{}^2 - k^2{}_0}}{k_0} = S^2{}_0 \tan\left(k_0 h\right), \qquad (2)$$

where, $S_0 = [(D-A) / D]^{0.5} \times \text{sinc}\,(\beta_{SSPP} \times (D - A) / 2)$ and $k_0 = \omega\sqrt{\varepsilon_{r\text{-}SSPP}} / c$. In (2), $\beta_{SSPP}$ is the phase constant of the SSPPs and $k_0 = \omega / c$. As the perfect electric conductor is substituted with a metal (gold here), the SSPP Ohmic attenuations are considered using the formulation in [19]. In this manner, SSPP propagation constant $\gamma_{SSPP} = \alpha_{SSPP} + j \times \beta_{SSPP}$ can be obtained, where $\alpha_{SSPP}$ is the attenuation constant. Similarly, the characteristic impedance of this transmission line is selected equal to the SSPP fundamental mode wave impedance: $Z_{SSPP} = \gamma_{SSPP} / (j\omega \times \varepsilon_{SSPP})$ where $\varepsilon_{SSPP} = \varepsilon_0 \times \varepsilon_{r\text{-}SSPP}$. Here, A = 20 μm, H = 70 μm, D= 60 μm have been considered in the design. The thickness of the metallic surface "t" is assumed to be larger compared to the skin depth in the interested frequency range.

*C. Details of the full-wave simulation and the analytical model*

The electromagnetic modeling of the proposed structure is performed using two different approaches, namely a transmission line (TL) formalism and a full wave simulator [20].

The transition of the electromagnetic wave from the SSPP on the grooved metallic surface to the 2D plasmons on the graphene layer can be characterized by cascading two TLs depicted in Fig. 1 (b). To this end, a section of the indented metallic surface with length $l_1$, characteristic impedance $Z_{SSPP}$ and propagation constant $\gamma_{SSPP}$ is considered as the first TL. The other sections of the SSPP waveguide and the exciting field are represented as a voltage source with internal resistance $Z_{SSPP}$. Furthermore, a small portion of the suspended graphene sheet adjacent to the metallic edge, with length $l_2$ is represented as the second TL with characteristic impedance $Z_G$ and propagation constant $\gamma_G$. The remaining part of the graphene layer is recognized as a load with impedance $Z_G$. In this manner, the scattering parameters of the equivalent circuit in Fig. 1 (a), calculated at reference planes 1 and 2, may be obtained using TL theory. This method provides a fast solution of the mentioned problem. However, it cannot include the effects of higher order modes which exist in the vicinity of the junction of the waveguides. These evanescent higher order modes exist near the discontinuity because of different characteristic impedances and propagation constants of the TLs. The evanescent modes specifically cause higher than expected attenuations, due to impedance mismatch.

For comparison and to provide more accurate results, a complete solution of Maxwell equation is performed using the numerical solver [20]. The simulation domain is excited by applying a wave port at x = 0 plane. In order to calculate scattering parameters, the methodology proposed in [16] is followed.

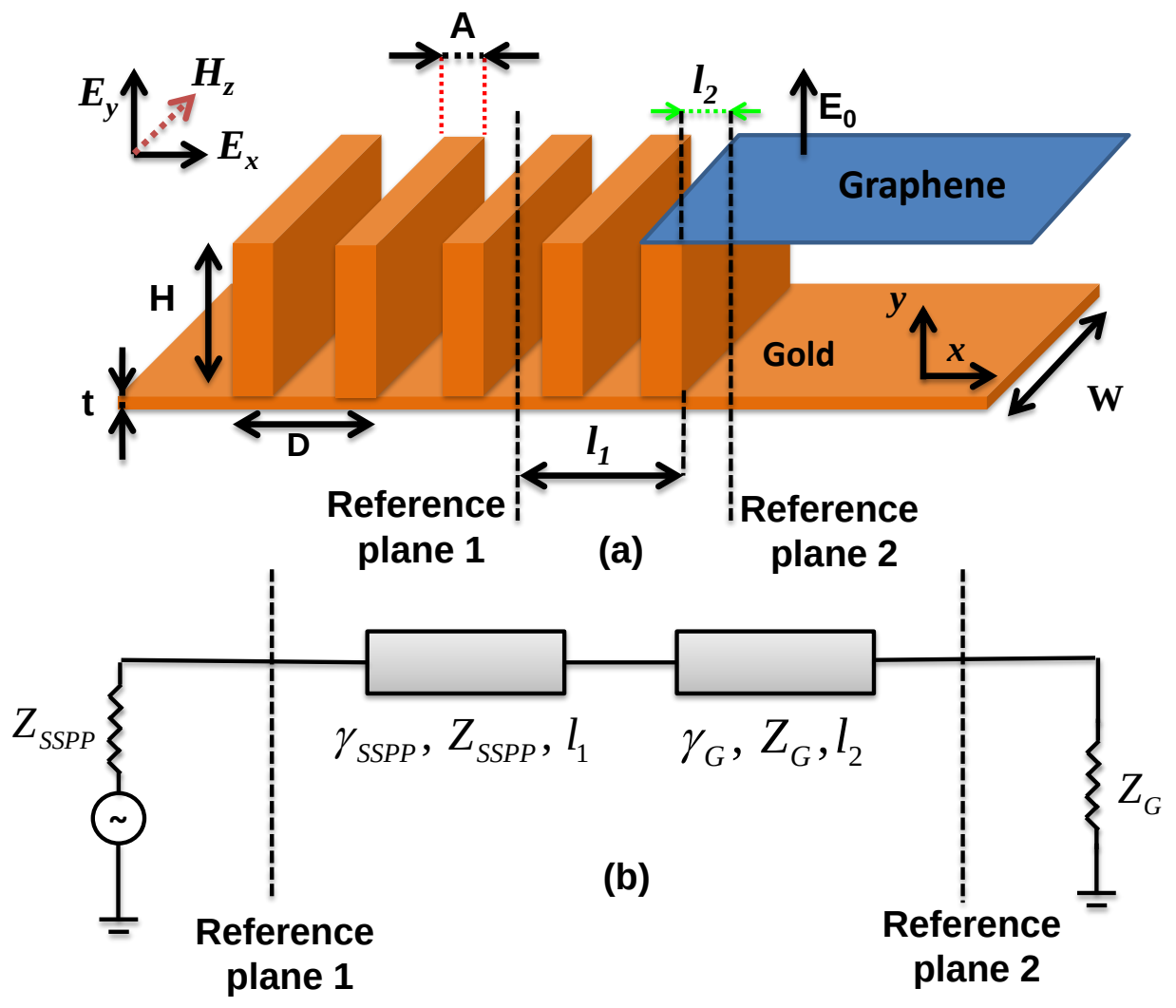

Fig. 1. (a) Proposed mechanism to excite 2D plasmons of a graphene sheet using SSPPs along a periodically indented metallic surface.

## III. RESULTS AND DISSCUSSION

In Fig. 2, the phase constants of the SSPPs and the 2D plasmons of the graphene sheet with various applied electric fields $E_0$ are depicted. As depicted in Fig. 2, the plasmons along both structures obtain higher momentums compared to the radiating mode counterpart ($k_0$), especially as f > 500 GHz. Moreover, slight changes in the chemical potential of the graphene lead to considerable variations in the 2D plasmon phase constants. It is also observed that there exists a single frequency for each Fermi energy, which the properties of the 2D plasmons are exactly similar to the ones of the SSPPs. It is expected that the transition of EM field from the SSPP to the plasmons on graphene can be ideally occurred at this frequency. Moreover, the differences between the phase constants of the 2D plasmons with $E_F$ = 0.18 and 0.16 eV, and the properties of the SSPPs are not very not very deep. Therefore, it is anticipated that an acceptable level of impedance matching between these waveguides may exist for these specific chemical potentials of the graphene.

Fig. 3 and Fig. 4 depict the calculated transmission ($S_{21}$) and input reflection ($S_{11}$) of the plasmonic waves. As shown in Fig. 3, the EM energy is transferred from the SSPPs to 2D plasmons with acceptable level of attenuation. Additionally, this technique is effective in a wide frequency range. Considering the input reflection coefficient in Fig. 4, it is understood that the matching between the transmission lines may be optimized in a certain frequency by changing the graphene chemical potential. The frequency of the minimum reflection coefficient in Fig. 4 is identical to the crossing point of the dispersion relations of the 2D plamsons and SSPPs.

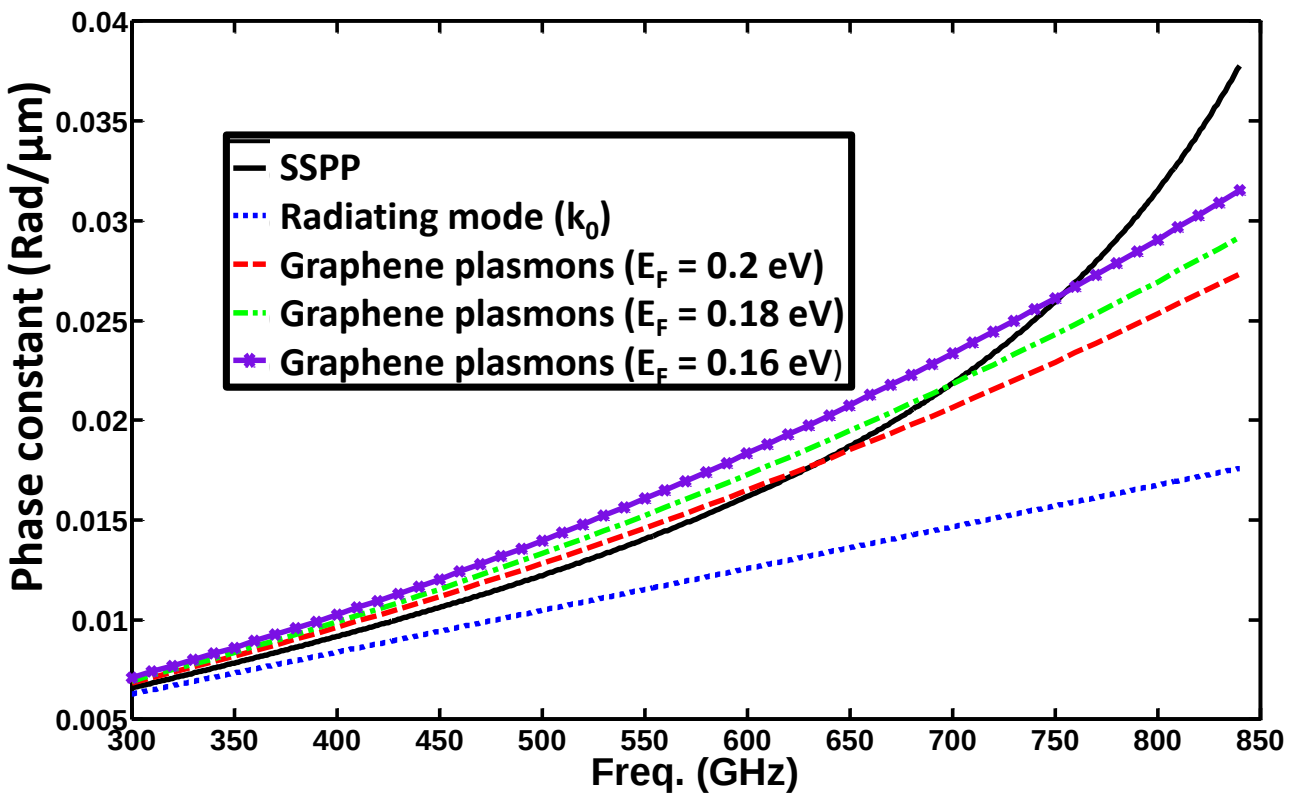

Fig. 2. Dispersion relation of the plasmons along the periodically corrugated metal and a suspended graphene with different Fermi energy levels obtained by the analytical model.

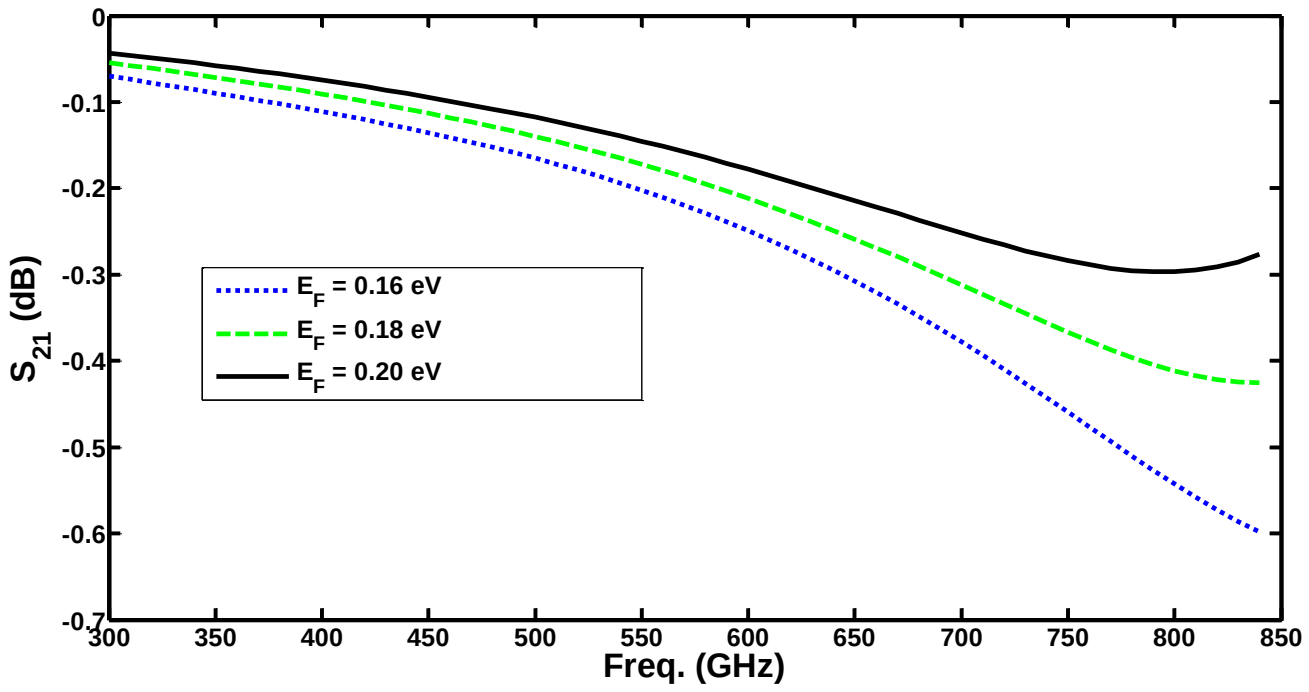

Fig. 3. Wave transmission from the corrogated metal to the suspended graphene with different Fermi energy levels.

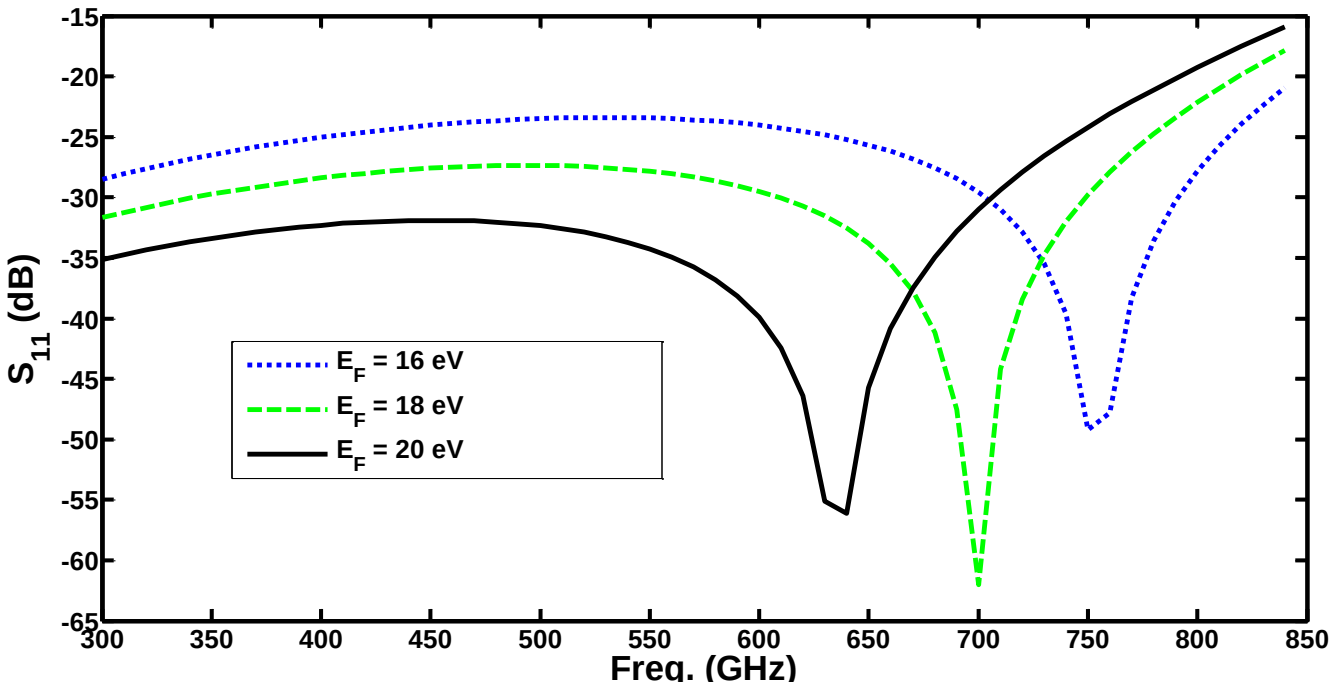

Fig. 4. Input wave reflection as transiting from the metallic surface to the graphene sheet with different Fermi energy levels.

Fig. 5 presents a comparison between $S_{11}$ and $S_{21}$ of the structure in Fig. 1. (a) (as $E_F$ = 0.16 eV) obtained by the analytical and numerical models. As depicted, there is a good similarity between estimated transmission using different approaches. However, HFSS predict slightly higher attenuations throughout the simulated frequency range which is due to the consideration of higher order modes which are not included into the analytical model. Favorably low $S_{11}$ reported by both methods verify the usefulness of the proposed technique to launch 2D plasmons in a wide frequency range.

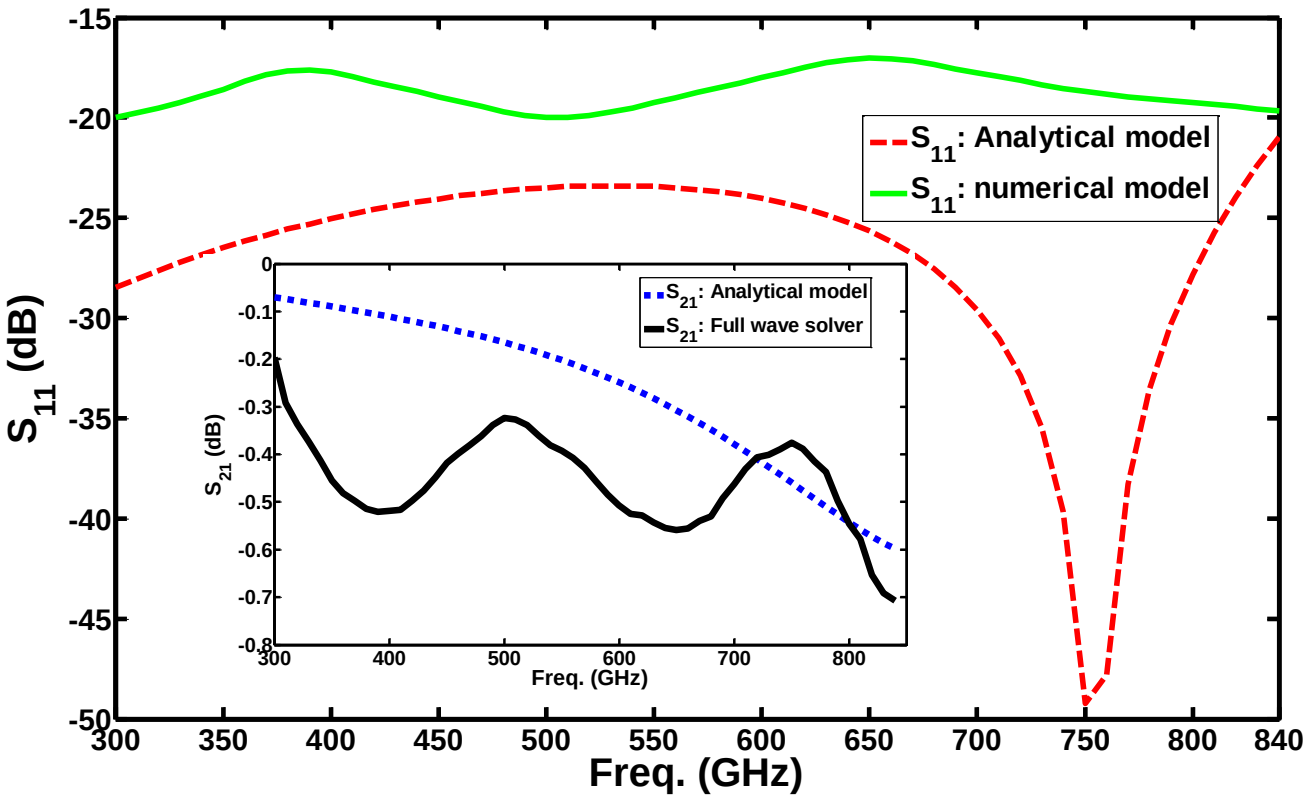

Fig. 5. A comparison between the reflection and transmission coefficients (as $E_F$ = 0.16 eV) calculated by the TL model and numerical solver.

## IV. Conclusion

In this paper, a promising technique to effectively launch 2D plasmons in a suspended graphene layer using spoof surface plasmons polaritons along a periodically indented metal is proposed. To show the effectiveness of the suggested method, an analytical technique based on transmission line theory is proposed. To verify the results, a full-wave commercial solver is employed.